\documentclass[aps,pre,amsmath,twocolumn,showpacs]{revtex4}
\usepackage{amsmath}
\usepackage{amsfonts}
\usepackage{graphicx}
\usepackage{hyperref}
\usepackage{mathtools}
\usepackage{verbatim}
\usepackage{epstopdf}
\usepackage{subfigure}
\usepackage{latexsym}
\usepackage{dcolumn}
\usepackage{epsf}
\usepackage{float}
\DeclarePairedDelimiter\abs{\lvert}{\rvert}

\usepackage{color}

\begin{document}
\title{Synchronization transition in Sakaguchi-Kuramoto model on complex networks with partial degree-frequency correlation}
\author{Prosenjit Kundu and Pinaki Pal}
\affiliation{Department of Mathematics, National Institute of Technology, Durgapur 713209, India}
\date{\today}
\begin{abstract}
{We investigate transition to synchronization in Sakaguchi-Kuramoto (SK) model on complex networks  analytically as well as numerically. Natural frequencies of a percentage ($f$) of higher degree nodes of the network are assumed to be correlated with their degrees and that of the remaining nodes are drawn from some standard distribution namely Lorenz distribution. The effects of variation of $f$ and phase frustration parameter $\alpha$ on transition to synchronization are investigated in detail. Self-consistent equations involving critical coupling strength ($\lambda_c$) and group angular velocity ($\Omega_c$) at the onset of synchronization have been derived analytically in the thermodynamic limit. For the detailed investigation we considered SK model on scale-free as well as Erd\H{o}s-R\'{e}nyi (ER) networks. Interestingly explosive synchronization (ES) has been observed in both the networks for different ranges of values of $\alpha$ and $f$. For scale-free networks, as the value of $f$ is set within $10\% \leq f \leq 70\%$, the range of the values of $\alpha$ for existence of the ES is greatly enhanced compared to the fully degree-frequency correlated case. On the other hand, for random networks, ES observed in a narrow window of $\alpha$ when the value of $f$ is taken within $30\% \leq f \leq 50\%$. In all the cases critical coupling strengths for transition to synchronization computed from the analytically derived self-consistent equations show a very good agreement with the numerical results.}     
\end{abstract}
 \pacs {05.45.Xt, 05.45.Gg, 89.75.Fb}
 \maketitle
\section{Introduction}
Kuramoto model (KM) has been widely used by the researchers for investigating the phenomenon of synchronization in weakly coupled oscillators due to its analytical accessibility and ability to capture essential features of synchronization~\cite{Kuramoto,acebrone:rmp77_2005,rodrigues:pr_2016}. Through a detailed analysis, Kuramoto~\cite{Kuramoto} established that the dynamics of the phases of a set of $N$ weakly coupled oscillators can be described by
\begin{equation}
\dot{\theta_i} = \omega_i + \sum_{j=1}^N F(\theta_j - \theta_i)~~(i = 1\dots N),
\end{equation}
where $\theta_i$ and $\omega_i$ are the phase and natural frequency of the $i$th oscillator and $F$ is a $2\pi$-periodic function, called coupling function. Since it's introduction, Kuramoto model has been employed with simple coupling function $F(\theta) = (\lambda/N)\sin{\theta}$, where $\lambda$ is the coupling strength, for theoretical understanding of synchronization in variety of systems appearing in physics, biology, and even in sociology~\cite{Pikovsky,Boccaletti-physrep,Arenas-physrep,buck:qrb_1988,neda:pre61_2000,yamaguchi:sc302_2003,wissenfeld:pre_1998,eckhardt:pre75_2007}. 
Synchronization has been studied under the classic paradigm of KM in all-to-all coupled global network in great detail considering unimodal distribution of the natural frequencies and second order transition to synchrony has been reported~\cite{Boccaletti-physrep,Arenas-physrep}. Researchers also considered different variations of KM model to investigate the effect of time delay~\cite{strogatz:prl1999,peron:pre_a2012}, network structure~\cite{net_structure}, non-trivial natural frequency distribution~\cite{Omelchenko} and several other factors on transition to synchronization~\cite{rodrigues:pr_2016}. 

Recently in an interesting variation of KM, Jesus et al.~\cite{jesus:2011} has considered degree-frequency correlated complex network of phase oscillators and reported first order transition to synchronization (Explosive Synchronization (ES)) for the first time. ES immediately has drawn the attention of the researchers and inspired a series of works. It has been reported in degree-frequency correlated networks of R\"ossler oscillators~\cite{leyva:prl108_2012}, second order Kuramoto oscillators~\cite{p_ji:PRL2013} and multiplex networks~\cite{Nicosia:prl118_2017}. On the other hand ES has also been reported in networks of oscillators where degree-frequency are not correlated~\cite{Xu:Sc_rep2016,Zhang:PRL2015,Leyva:PRE2013}. The role of degree-frequency correlation in network synchronization has been investigated in~\cite{skardal:epl101}. For a detailed review on explosive synchronization, please see~\cite{bocaletti:PR2018_660} and references there in. 

Notably, in a recent work Peron and Rodrigues~\cite{Peron:pre86_2012} determined analytical expression of coupling strength for transition to synchronization including ES using mean field approach proposed by Ichinomia~\cite{ichinomiya:pre_2004}. Following similar mean field approach,  Coutinho et al.~\cite{Coutinho:pre87_2013} determined self consistent equations for analyzing transition to synchronization which has been successfully employed afterwards in studying disorder induced ES~\cite{pre:skardal2014,Zhang:PRE2013}. Interestingly, it has been discovered by Pinto and Saa~\cite{Pinto_saa2015} that ES is enhanced in scale free network with partial degree-frequency correlation. 

Although, ES has been investigated in Kuramoto model by several researchers, it is not investigated in that much detail in Sakaguchi-Kuramoto (SK) model~\cite{Sakaguchi}. Kundu et al.~\cite{Kundu2017} recently exploited the method proposed by Coutinho et al.~\cite{Coutinho:pre87_2013} based on mean field approximation  to derive self-consistent equations to investigate transition to synchronization in degree-frequency correlated SK model on complex networks and determined the effect of phase frustration parameter both on first and second order transition to synchronization.

In this paper, we consider SK model on complex networks with partial degree frequency correlation i.e. a percentage ($f$) of higher degree nodes are degree-frequency correlated and natural frequencies of the remaining nodes are drawn from a standard distribution. We perform analytical as well as numerical investigation of the SK model on complex networks to understand the effect of variation of the values of $f$ and phase frustration parameter $\alpha$ on transition to synchronization including ES which is not done earlier. 

\section{Mean Field Approach}
We consider a complex network of $N$ coupled phase oscillators (Sakaguchi-Kurmaoto model~\cite{Sakaguchi})   
 \begin{eqnarray}\label{eqn1}
\frac{d\theta_i}{dt} &=& \omega_i +\lambda\sum_{j=1}^{N} A_{ij}\sin(\theta_j - \theta_i-\alpha), ~i = 1\dots N,
\end{eqnarray}
with phase $\theta_i$ and natural frequency $\omega_i$ of the $i$th oscillator. The structure of the complex network is given by the coupling matrix $A = (A_{ij})_{N\times N}$ with $A_{ij} = 1$ if $i$th and $j$th oscillators are connected and $A_{ij} = 0$ otherwise, $\alpha\in[0,\frac{\pi}{2})$ is the phase-lag parameter and $\lambda$ is the coupling strength. 
The degree of synchronization in the network is measured by the order parameter $r$ given by
 \begin{eqnarray}
 r e^{i\psi} = \frac{\sum_{j=1}^{N} k_j e^{i\theta_j}}{\sum_{j=1}^{N}k_j},\label{r}
\end{eqnarray}
where $\psi$ is the average phase of the ensemble at time $t$ and $k_j$ is the degree of the $j$th node of the network. The value of $r$ ranges from $0$ (incoherent state) to $1$ (fully synchronized state). 

We now assume that the vertices of the network with degree greater than a certain threshold value $k_{*}$ are degree-frequency correlated i.e. $\omega_i = k_i~(k_i\geq k_{*})$ and the natural frequencies of other vertices ($(k_i < k_{*}$) are drawn from a distribution $g(\omega)$. So we write the joint probability distribution for a vertex of the network with degree $k$ and natural frequency $\omega$ as~\cite{Pinto_saa2015}
 \begin{eqnarray}\label{g_om_k}
G(k,\omega)&=& [\delta (\omega -k)P(k)-g(\omega)P(k)]H(k-k_{*})\nonumber \\ && + g(\omega)P(k), 
\end{eqnarray}
where $\delta$, $H$ and $P$ represent the Dirac delta function, Heaviside step function and degree distribution function of the network respectively. We can say here that the natural frequencies of the oscillators are drawn from the distribution $G(k,\omega)$. The following may now be easily checked
\begin{eqnarray}
\int G(k,\omega)d\omega &=& P(k),\\ \label{p_k}
\int G(k,\omega)dk &=& P(\omega)H(\omega -k_{*}) + \beta_{1} g(\omega),\label{g_om_k1}
\end{eqnarray}
where $\beta_{1} =\int_{k_{min}}^{k_{*}} P(k)dk$ and $k_{min}$ is the network minimum degree. We also note that the average degree $\langle k \rangle$ of the network is given by
 \begin{eqnarray}\label{mean_k}
\langle k \rangle=\int k dk \int d\omega G(k,\omega)=\int_{k_{min}}^{\infty} kP(k)dk.
\end{eqnarray} 

Now following Ichinomiya \cite{ichinomiya:pre_2004} we assume that the distribution density of oscillators of vertices with phase $\theta$ at time $t$ with given degree $k$ and frequency $\omega$ is given by the function $\rho (k,\omega,\theta,t)$ and it is normalized as 
\begin{eqnarray}
\int_{0}^{2\pi} \rho (k,\omega,\theta,t) d\theta =1.
\end{eqnarray}

Therefore, in the continuum limit ($N \rightarrow \infty$), equation~(\ref{eqn1}) can be written as 
\begin{eqnarray}
\frac{d\theta(t)}{dt} & = & \omega  + \frac{\lambda k}{\langle k \rangle} \int d\omega' \int dk' \int d\theta' k' G(k',\omega')  \nonumber\\
&&\times \rho(k',\omega',\theta',t) \sin(\theta' -\theta -\alpha)\label{eqn12}
\end{eqnarray}
and the order parameter is given by
\begin{eqnarray}
r e^{i\psi} & = & \frac{1}{\langle k\rangle} \int d\omega \int dk \int d\theta k G(k,\omega)\rho(k,\omega, \theta,t) e^{i\theta}.\nonumber\\\label{r_cont}
\end{eqnarray}
The conservation of the oscillators of the network gives the equation of continuity
\begin{equation}
\frac{\partial \rho}{\partial t} + \frac{\partial}{\partial \theta}(\rho v) = 0,
\label{continuity}
\end{equation}
for the density function $\rho$, where $v$ is the right hand side of the equation (\ref{eqn12}).

From equations (\ref{r_cont}) and (\ref{eqn12}) we get

 \begin{eqnarray}
\frac{d\theta}{dt} & = & 
\omega + \lambda k r \sin(\psi -\theta -\alpha).
\label{eqn15}
\end{eqnarray}
We write average phase $\psi(t) = \Omega t$ where $\Omega$ is the group angular velocity and introduce a new variable $\phi$ with $\phi(t)=\theta(t)-\psi(t)+\alpha$. In terms of this new variable, equation (\ref{eqn15}) can be written as 
\begin{eqnarray}
\frac{d\phi}{dt} & = & 
\omega- \Omega - \lambda k r \sin(\phi)\label{eqn6}
\end{eqnarray}
and the equation of continuity (\ref{continuity}) takes the form
\begin{eqnarray}
 \frac{\partial}{\partial t} \rho(k,\omega, \phi,t) + \frac{\partial}{\partial \phi} [v_\phi \rho(k,\omega, \phi,t)]=0, \label{eqn_cont}
 \end{eqnarray}
where $v_\phi = \frac{d\phi}{dt}$. In the steady state, we have $\frac{\partial}{\partial t} \rho(k,\phi,t) =0$. 
\noindent Therefore, steady state solution for the density function $\rho$ is given by
\begin{eqnarray}
 \rho(k,\omega, \phi)=\begin{cases}
\delta \left(\phi-arc\sin{\left(\frac{\omega-\Omega}{k\lambda r}\right)}\right), & \abs*{\frac{\omega-\Omega}{k\lambda r}} \leq 1 \\
\frac{C_1(k,\omega)}{\omega-\Omega -k\lambda r \sin(\phi)}, & \abs*{\frac{\omega-\Omega}{k\lambda r}} > 1,
\end{cases}
\end{eqnarray} 
 where $C_1(k,\omega)=\frac{\sqrt{(\omega-\Omega)^2 -(k\lambda r)^2 }}{2\pi}$ is the normalization constant.The first part corresponds to the locked oscillators and the second part is associated with drift oscillators of the network.
 
Hence the order parameter can be rewritten as 
\begin{eqnarray}
r &=& \frac{1}{\langle k\rangle} \int_{k_{min}}^{\infty} k dk \bigg[\int_{\Omega-\lambda r k}^{\Omega+\lambda r k} d\omega G(k,\omega) \nonumber\\
&& \times \exp {i(\arcsin (\frac{\omega -\Omega}{\lambda k r})-\alpha)}\nonumber\\
&+&\int_{\Omega+\lambda r k}^{\infty} d\omega G(k,\omega) \int d\phi \frac{c_1(\omega,k)}{\omega -\Omega - \lambda r k\sin \phi} e^{i(\phi-\alpha)} \nonumber\\
&& + \int_{-\infty}^{\Omega-\lambda r k} d\omega G(k,\omega) \int d\phi \frac{c_1(\omega,k)}{\Omega-\omega + \lambda r k} e^{i(\phi-\alpha)}\bigg] \nonumber\\ \label{r_d_l}
 &=& \frac{1}{\langle k\rangle}\bigg[I_1(\lambda r) + I_2(\lambda r)\bigg] ~(\mathrm{say}). 
 \end{eqnarray}
Here $I_1(\lambda r)$ is the first part of the above integral which gives the contribution of locked oscillators to the order parameter, while $I_2(\lambda r)$ is the sum of second and third parts of the integral giving the contribution of drift oscillators to the order parameter.  
 
Now to evaluate the integral  
 \begin{eqnarray}
I_1(\lambda r) &=& e^{-i\alpha}\int_{k_{min}}^{\infty} k dk \int_{\Omega-\lambda r k}^{\Omega+\lambda r k} d\omega G(k,\omega)\nonumber\\ 
&& \times \exp {(i\arcsin (\frac{\omega -\Omega}{\lambda k r}))},
 \label{I2}
 \end{eqnarray}
we introduce a new variable $\eta = \frac{\omega - \Omega}{\lambda kr}$ and the integral reduces to
\begin{eqnarray}
I_1(\lambda r) &=& e^{-i\alpha} \lambda r\int_{-1}^{1} d\eta \exp {(i \arcsin (\eta))}  \nonumber \\
&&  \times \int_{k_{min}}^{\infty} k^2 dk G(k,\Omega+\lambda r k \eta).
\label{I_2_1}
\end{eqnarray}

Then using~(\ref{g_om_k}) we have 
\begin{eqnarray}
 \int_{k_{min}}^{\infty} k^2 dk G(k,&\Omega& + \lambda r k \eta ) = \int_{k_{min}}^{k_*} k^2 g(\Omega+ \lambda k r \eta) P(k) dk  \nonumber \\
&& + \int_{k_*}^{\infty} k^2 \delta (\Omega+ \lambda k r \eta -k)P(k)dk,
 \label{I_2_2}
 \end{eqnarray} 
 and hence
 \begin{eqnarray}
I_1(\lambda r) &=& e^{-i\alpha} \lambda r\int_{-1}^{1} d\eta \lbrace \sqrt{1-\eta^2} +i\eta \rbrace \nonumber \\
&& \times \bigg[ \int_{k_{min}}^{k_*} k^2 P(k) g(\Omega +\lambda r k \eta) dk \nonumber \\
&& +\frac{1}{\abs{1-\lambda r \eta}} \bigg(\frac{\Omega}{1-\lambda r \eta}\bigg)^2  \nonumber \\
&& \times P(\frac{\Omega}{1-\lambda r \eta})H(\frac{\Omega}{1-\lambda r \eta}-k_*)\bigg].
 \label{I_2_3}
 \end{eqnarray}

For drift oscillators we have,  
\begin{eqnarray}
 I_2(&\lambda r&)= e^{-i\alpha}\int_{k_{min}}^{\infty} k dk \bigg[\int_{\Omega+\lambda r k}^{\infty} d\omega G(k,\omega) \int d\phi \nonumber\\
 && \times \frac{c_1(\omega,k)e^{i\phi}}{\omega -\Omega - \lambda r k\sin \phi}  \nonumber\\
&& + \int_{-\infty}^{\Omega-\lambda r k} d\omega G(k,\omega) \int d\phi \frac{c_1(\omega,k)e^{i\phi}}{\Omega-\omega + \lambda r k \sin \phi}\bigg]. \nonumber \\
 \label{drift_1}
 \end{eqnarray}
Substituting $\omega=\Omega+\lambda k r \eta$ and $\omega=\Omega-\lambda k r \eta$ in the first and second term of the above integral respectively we get 
\begin{eqnarray}
I_2(\lambda r)&=& \lambda r \frac{e^{-i\alpha}}{2\pi}\int_{k_{min}}^{\infty} k^2 dk \int_{1}^{\infty} \sqrt{\eta^2-1} d\eta\times \int_{0}^{2\pi} d\phi \nonumber\\ 
&& \times \bigg[\frac{G(k,\Omega+\lambda r k\eta)}{\eta-\sin \phi} + \frac{G(k,\Omega-\lambda r k\eta)}{\eta+\sin \phi}\bigg]e^{i\phi}. \nonumber \\
 \label{drift_2}
\end{eqnarray}
Now it can be easily shown that  
\begin{eqnarray}
\int_{0}^{2\pi} \frac{e^{i\phi}}{\eta+\sin \phi}d\phi=-2\pi i\frac{\abs{\eta}-\sqrt{\eta^2 -1}}{\sqrt{\eta^2 -1}}.\label{int1}
\end{eqnarray}
Using~(\ref{int1}) in equation (\ref{drift_2}) we get 
\begin{eqnarray}
I_2(\lambda r)&=& \lambda r i e^{-i\alpha} \int_{k_{min}}^{\infty} k^2 dk \int_{1}^{\infty} f(\eta)d\eta \nonumber\\ 
&& \times\bigg[{G(k,\Omega+\lambda r k\eta)} - {G(k,\Omega-\lambda r k\eta)}\bigg],\nonumber \\
 \label{drift_3}
\end{eqnarray}
where $f(\eta)=\eta-\sqrt{\eta^2-1}$.

We can further write 
$$I_2(\lambda r)=I_2^{(a)}(\lambda r)+I_2^{(b)}(\lambda r),$$
where, 
\begin{eqnarray}
I_2^{(a)}(\lambda r)&=& \lambda r i e^{-i\alpha} \bigg[\int_{max(k_*,\frac{\Omega}{1-\lambda r})}^{\infty} dk k P(k)\frac{1}{\lambda r}f(\frac{k-\Omega}{\lambda k r}) \nonumber \\
&& -\int_{k_*}^{max(k_*,\frac{\Omega}{1-\lambda r})} dk k P(k)\frac{1}{\lambda r}f(\frac{\Omega -k}{\lambda k r})\bigg],
 \label{drift_4}
\end{eqnarray}
\begin{eqnarray}
I_2^{(b)}(\lambda r)&=&  \lambda r i e^{-i\alpha} \int_{k_{min}}^{k_*} dk k^2 P(k)\int_{1}^{\infty} d\eta f(\eta) \times \nonumber \\
&& \bigg[{g(\Omega+\lambda r k\eta)} - {g(\Omega-\lambda r k\eta)}\bigg],
 \label{drift_5}
\end{eqnarray}
using the definition of $G(k, \omega)$.\\

Finally, to determine the critical coupling strength ($\lambda_c$) and group angular velocity ($\Omega_c$) for the onset of synchronization, we consider the limit $r \rightarrow 0^+$ in the above expressions of $I_1(\lambda r)$, $I_2^{(a)}(\lambda r)$ and $I_2^{(b)}(\lambda r)$ and get

\begin{eqnarray}
\langle k \rangle &=& p.v.\bigg[\lambda_c(\sin \alpha +i\cos \alpha) \bigg\lbrace \int_{k_*}^{\infty} dk  P(k)\frac{k^2}{k-\Omega_c} \nonumber \\ &&+ \int_{-\infty}^{\infty} d\omega\frac{g(\omega)}{k-\Omega_c}\bigg]  +\lambda_c(\cos \alpha -i\sin \alpha )\nonumber \\ 
&&\int_{-1}^{1} \lbrace \sqrt{1-\eta^2} +i\eta \rbrace d\eta \nonumber \\ && \times \bigg[ g(\Omega_c)\beta_2 +\Omega_c^2 P(\Omega_c)H(\Omega_c -k_*)\bigg],
 \label{r_k}
\end{eqnarray}
where 
\begin{eqnarray}
\beta_2=\int_{k_{min}}^{k_*} k^2 P(k) dk,
\label{beta2}
 \end{eqnarray}
$\Omega_c$, $\lambda_c$ are the values of $\Omega$ and $\lambda$ respectively in the limit $r\rightarrow 0^+$ and the abbreviation $p.v.$ stands for principal value. Note that for the evaluation of the integrals we have used the approximation 
 \begin{eqnarray}
\frac{1}{\lambda r}f(\frac{\abs{k-\Omega}}{\lambda k r}) \approx\frac{1}{2} \abs{\frac{k}{k-\Omega}}.
 \label{drift_6}
\end{eqnarray}
 
Now comparing the real and imaginary parts we get,
\begin{eqnarray}
\langle k \rangle &=&  \lambda_c \times \bigg[ p.v. \bigg\lbrace \frac{\sin \alpha}{2} \int_{k_*}^{\infty} dk  P(k)\frac{k^2}{k-\Omega_c}\bigg\rbrace \nonumber \\ && + p.v.\bigg\lbrace \frac{\beta_2 \sin \alpha }{2}\int_{-\infty}^{\infty} d\omega \frac{g(\omega)}{\omega-\Omega_c}\bigg\rbrace \nonumber \\ && + \bigg\lbrace{ P(\Omega_c) \Omega_c^2  H(\Omega_c-k_*) + \beta_2 g(\Omega_c)} \bigg\rbrace \nonumber \\
&& \times \cos \alpha\int_{-1}^{1} \sqrt{1-\eta^2} d\eta \bigg]
 \label{real_r} 
\end{eqnarray}
and 
\begin{eqnarray}
0 &=& \bigg[ p.v. \bigg\lbrace \frac{\cos \alpha}{2} \int_{k_*}^{\infty} dk  P(k)\frac{k^2}{k-\Omega_c}\bigg\rbrace \nonumber \\ && + p.v.\bigg\lbrace \frac{\beta_2 \cos \alpha }{2}\int_{-\infty}^{\infty} d\omega \frac{g(\omega)}{\omega-\Omega_c}\bigg\rbrace \nonumber \\ && - \bigg\lbrace{ P(\Omega_c) \Omega_c^2  H(\Omega_c-k_*) + \beta_2 g(\Omega_c)} \bigg\rbrace \nonumber \\
&& \times \sin \alpha\int_{-1}^{1} \sqrt{1-\eta^2} d\eta \bigg].
 \label{im_r}
\end{eqnarray}
Combining Eq.(\ref{real_r}) and Eq.(\ref{im_r}) we get 
\begin{eqnarray}
\lambda_c &=& \frac{2 \cos \alpha \langle k \rangle}{\pi \lbrace{ P(\Omega_c) \Omega_c^2  H(\Omega_c-k_*) + \beta_2 g(\Omega_c)} \rbrace},
 \label{lambda_c2} 
\end{eqnarray}
where $\Omega_c$ can be found from the equation
\begin{eqnarray}
\bigg\lbrace P(\Omega_c)&\Omega_c^2& H(\Omega_c-k_*) + \beta_2 g(\Omega_c)\bigg\rbrace \pi \tan \alpha \nonumber \\
 &=& \bigg[ p.v. \bigg\lbrace \int_{k_*}^{\infty} dk  P(k)\frac{k^2}{k-\Omega_c}\bigg\rbrace \nonumber \\ &+& p.v.\bigg\lbrace \beta_2\int_{-\infty}^{\infty} d\omega \frac{g(\omega)}{\omega-\Omega_c}\bigg\rbrace \bigg].
 \label{Omega_c}
\end{eqnarray}
Therefore, from the self-consistent equations (\ref{lambda_c2}) and (\ref{Omega_c}) we can now determine critical coupling strength ($\lambda_c$) and group angular frequency ($\Omega_c$) for the onset of synchronization for a given network, $k_{*}$ and $\alpha$. In the next section, we illustrate the theory developed in this section by performing numerical simulations with scale-free and random networks.

\section{Numerical simulation results}
For illustration of the theory developed in the previous section to determine critical coupling strength for the onset of synchronization we consider scale-free as well as Erd\"{o}s-R\'{e}nyi networks. We then numerically integrate the networks of oscillators using fourth order Runge-Kutta (RK4) scheme with time step $\Delta t = 0.001$. Integration is performed for $5000$ time unit in each case and out of which the data upto $2000$ time unit is removed as transient. Rest of the data is used for computation of various quantities. Note that to determine the type of transition to synchrony, we need to continue the integration both in forward and backward direction. For forward integration, we start with a small value of $\lambda$ and increase the value of $\lambda$ with step size $\Delta{\lambda} = 0.02$. Random initial condition is used for the first integration and for subsequent integration, the last point of the previous integration is used as initial condition. During backward continuation, we start with the highest value of $\lambda$ and decrease the value $\lambda$ with step size $\Delta{\lambda} = 0.02$. 
\subsection{Scale-free network}
\begin{figure}
\includegraphics[height=0.28\textheight,width=0.48\textwidth]{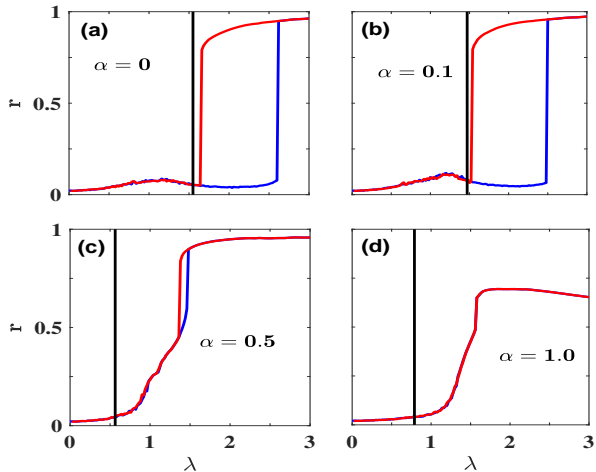}
\caption{Effect of $\alpha$ on transition to synchronization for a scale-free network of size $N = 2000$, degree distribution exponent $\gamma = 2.9$ and mean degree $\langle k \rangle = 8$ where $10\%$ of the higher degree nodes of the network are degree-frequency correlated. Numerically computed order parameter ($r$) is plotted as a function coupling strength  $(\lambda)$ for forward (solid blue curve) and backward (solid red curve) continuation for four values of $\alpha$.  The vertical black line indicates the critical coupling strength $(\lambda_c)$ calculated from the self consistent equations (\ref{lambda_c2}) and (\ref{Omega_c}). } \label{10prcnt}
\end{figure} 
\begin{figure}
     \includegraphics[height=0.28\textheight, width=0.48\textwidth]{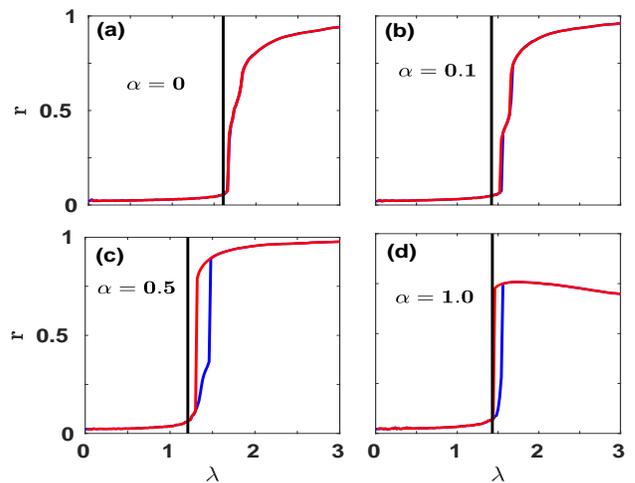}
     \caption{Numerically computed order parameter for SK model on a scale-free network of size $N = 2000$, $\gamma = 2.9$ with $50\%$ degree-frequency correlated nodes for four values of $\alpha$. Blue and red curves represent the order parameters computed for forward and backward continuation respectively.  Critical coupling strength ($\lambda_c$) calculated from the self-consistent equations are shown with vertical black lines for four values of $\alpha$.}  \label{50prcnt}
\end{figure}

We start with a degree-degree uncorrelated~\cite{Newmaan:2002} scale-free network of size $N = 2000$ with degree distribution exponent $\gamma=2.9$ and mean degree $\langle k \rangle = 8$. $10\%$ of the higher degree nodes of the network are taken to be degree-frequency correlated and the natural frequencies of the remaining $90\%$ of the nodes are drawn from Lorenzian distribution of $0$ mean (i.e. $g(\omega)=\frac{\Delta}{\pi(\Delta^2 + \omega^2)}$, $\Delta$ is the half width and in this paper we take $\Delta = 5$). The value of $k_*=12$ in this case. The network is then simulated for four different values of $\alpha$ and the order parameter is calculated both for forward and backward continuation as a function of the coupling strength $\lambda$. The numerically computed order parameters for four values of $\alpha$ both for forward and backward continuation are shown in figure~\ref{10prcnt}. From the figure we observe that the system exhibits first order transition to synchronization (explosive synchronization (ES)) both in presence and absence of phase frustration ($\alpha$) even though the percentage of degree-frequency correlated nodes is very less ($10\%$). It is observed that as the value of $\alpha$ is increased, the width of the hysteresis loop is decreased and ES ceased to exist for $\alpha = 0.95$. The transition to synchronization is found to be of second order for higher values of $\alpha$. 
 \begin{figure}
     \includegraphics[height=0.28\textheight, width=0.48\textwidth]{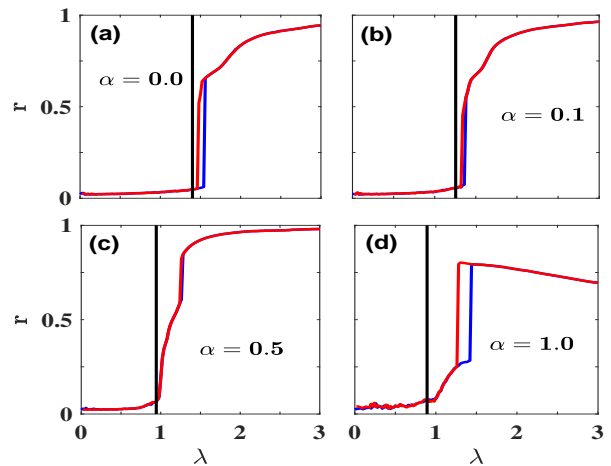}
     \caption{Effect of $\alpha$ on transition to synchronization for SK-model on the same scale-free network of size $N = 2000$ and $\gamma = 2.9$ when $70\%$ of oscillators are degree frequency correlated as obtained from numerical simulation. Critical coupling strengths computed from self-consistent equations are shown with vertical black lines.} 
\label{70prcnt}
\end{figure}  

We then calculate critical coupling strength $\lambda_c$ for the onset of synchronization during backward continuation using the self-consistent equations (\ref{lambda_c2}) and (\ref{Omega_c}). These critical coupling strengths are shown with vertical black lines in the figure~\ref{10prcnt}. In each case, the analytically computed critical coupling strength is found to be in good agreement with the numerical simulation result. It is apparent from the figure~\ref{10prcnt} that as the value of $\alpha$ increases, the value of the critical coupling strength $\lambda_c$ first decrease and then increases. 

Next we perform numerical simulation of the same scale-free network by setting $k_*=6$ which makes  nearly $50\%$ of the higher degree nodes degree-frequency correlated and the natural frequencies of the remaining $50\%$ nodes are drawn from Lorenz distribution of zero mean. The order parameter computed from the numerical simulation data for different values of $\alpha$ are shown in figure~\ref{50prcnt} both for forward and backward continuation. Interestingly, it is observed in figure~\ref{50prcnt}(a) that the transition to synchronization is second order in absence of phase frustration ($\alpha = 0$). Figures~\ref{50prcnt}(b)-(d) show that the transition to synchronization becomes first order (ES) from second order as the value of $\alpha$ is increased. 
We further observe from figure~\ref{50prcnt} that the transition to synchronization remains first order (ES) for higher values of $\alpha~(=1.0)$. In this case also we calculate the critical coupling strength $\lambda_c$ from the self-consistent equations and these are shown with vertical black lines in the figure~\ref{50prcnt} for four values of $\alpha$. From the figure we note that these critical coupling strengths match closely with the numerical simulation results.
\begin{figure}
     \includegraphics[height=!,width=0.48\textwidth]{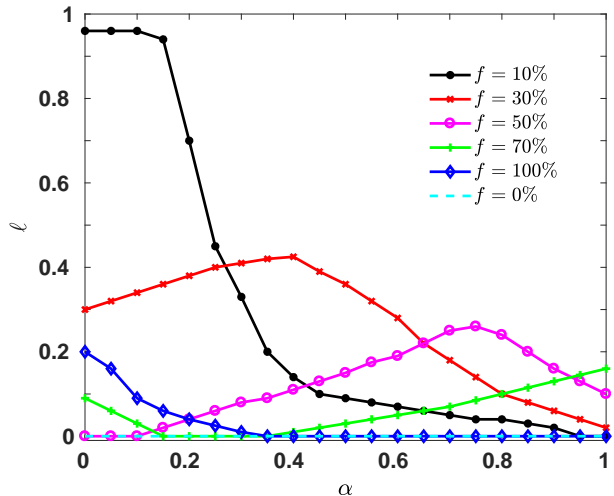}
     \caption{Width of hysteresis loop $\ell$ as a function of frustration parameter $\alpha$ for the SK-model on scale-free network of size $N=2000$ and degree distribution exponent $\gamma = 2.9$ for different values of $f$ as obtained from numerical simulation.}\label{hist_width}
\end{figure}

As the number of degree-frequency correlated oscillators of the SK-model on the same scale-free network of size $N = 2000$ is set to nearly $70\%$, ES with a smaller hysteresis loop is again observed in the absence of frustration (see figure~\ref{70prcnt}(a)). With increase of the value of phase frustration parameter $\alpha$, the hysteresis behavior is annihilated near $\alpha=0.3$ but surprisingly, ES returns with the further increment of $\alpha$. This behavior has been shown in the figures~\ref{70prcnt}(b)-(d). The critical coupling strengths ($\lambda_c$) computed from the self-consistent equations (\ref{lambda_c2}) and (\ref{Omega_c}) shown in figure~\ref{70prcnt} which closely match with the numerical results (see the vertical black lines in figure~\ref{70prcnt}). Note that if all the oscillators are degree-frequency correlated then for lower values of $\alpha$ system shows ES and for higher values of $\alpha$ second order transition to synchronization is observed, the details of which has already been reported in~\cite{Kundu2017}. On the other hand, if there is no correlation between the degree and the frequency of the oscillators, then only second order transition to synchronization is observed. 

It is evident from the previous discussion that the nature of transition to synchronization crucially depends upon the phase frustration parameter $\alpha$ and the percentage $f$ of the degree-frequency correlated higher degree nodes of the networks. For understanding this dependence in detail we numerically compute the width of the hysteresis loop $\ell$ as a function of $\alpha$ and $f$ when $\lambda$ is varied in the range $0$ to $3$ for the SK-model on the scale-free network of size $N = 2000$ and degree distribution exponent $\gamma = 2.9$. Figure~\ref{hist_width} shows the variation of the width of the hysteresis loop as a function of $\alpha$ for various values of $f$ ranging between $0\%$ and $100\%$. From the figure we observe that for $f = 0\%$ (cyan curve in the figure~\ref{hist_width}), the value of $\ell$ always remains zero indicating that there is no hysteresis in the entire range of $\alpha$ i.e. transition to synchronization is second order. For $f = 10\%$ we observe from the figure that widths of the hysteresis loops are quite large for small values of $\alpha$, while for larger values of $\alpha$, the width decreases and finally becomes zero near $\alpha =1$. So ES is observed for $f = 10\%$ in a large range of $\alpha$. ES is observe in the entire range of $\alpha$ for $f = 30\%$ (red curve in the figure~\ref{hist_width}). For higher percentage of degree-frequency correlation both first order and second order transition to synchronization are observed. For $f = 50\%$, second order transition is found for smaller values of $\alpha$, while for larger values of $\alpha$, ES synchronization is observed (see magenta curve in the figure~\ref{hist_width}). On the other hand, for $f = 70\%$, ES is observed for smaller and larger values of $\alpha$, while second order transition is observed for an intermediate range of $\alpha$ (green curve in the figure~\ref{hist_width}). It is interesting to note here that the range of $\alpha$ for the existence of ES is greatly enhanced when $10\%\leq f\leq 70\%$ compared to the one observed for $f = 100\%$ (black curve in the figure~\ref{hist_width}). 

\subsection{Erd\H{o}s-R\'{e}nyi network}
\begin{figure}
     \includegraphics[height=0.28\textheight, width=0.48\textwidth]{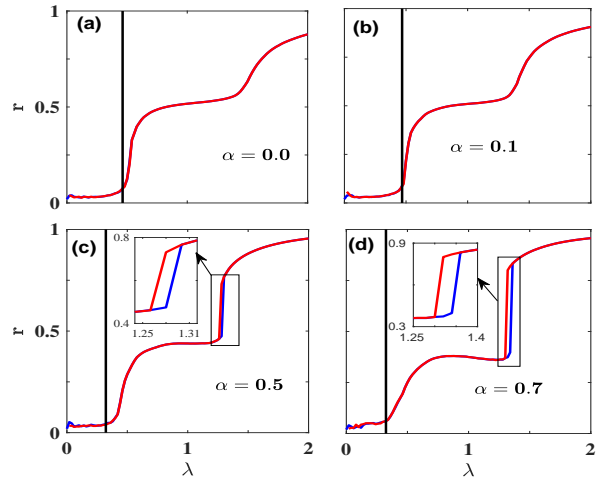}
     \caption{Effect of $\alpha$ on transition to synchronization for a ER network of size $N = 2000$ and mean degree $\langle k \rangle = 12$ where nearly $50\%$ of oscillators are degree frequency correlated. Numerically computed order parameter $r$ is plotted as a function of $\lambda$ for four values of $\alpha$. Critical coupling strengths calculated from the self-consistent equations are shown with vertical black lines. Zoomed views of the marked regions are shown in the insets.}  
 \label{ER50prcnt}
\end{figure}
We also consider here the SK-model on an Erd\H{o}s-R\'{e}nyi (ER) network of size $N=2000$ and mean degree $\langle k \rangle =12$ for the illustration the theory developed in the previous section. Numerically we simulate this ER network by varying the percentage of the degree-frequency correlated oscillators in the same range of $\alpha$ as was done in previous subsection and observe that the transition to synchronization generally is of second order type. However, for $f = 30\%$ and $f = 50\%$ we observe a narrow window of $\alpha$ where we observe ES of thin width of hysteresis loop. Figure~\ref{ER50prcnt} shows the numerically computed order parameter of the system for forward and backward continuation when $f = 50\%$. The figure clearly shows the existence of ES for $\alpha = 0.5$ and $0.7$ with very thin width of hysteresis loop. Note that for smaller and larger values of $\alpha$ only second order transition to synchronization is observed. For this network also we determine the critical coupling strength ($\lambda_c$) for the onset of synchronization using the self-consistent equations~(\ref{lambda_c2}) and (\ref{Omega_c}) and shown with vertical black lines in the figure~\ref{ER50prcnt}. It is observed that the critical coupling strength for the onset of synchronization matches closely with the numerical results.

\section{Conclusions}
In this paper, we have derived self-consistent equations for determining critical coupling strength for the onset of synchronization in partial degree-frequency correlated SK model on complex networks. In these networks, a percentage ($f$) of the higher degree nodes are assumed to be degree-frequency correlated and the natural frequencies of the remaining nodes are drawn from some standard distribution. The critical coupling strengths calculated from the self-consistent equations for different networks and other parameter values namely phase frustration parameter $\alpha$ and the percentage $f$ of degree-frequency correlated nodes of the network are found to match closely with the numerical simulation results. Moreover, we perform detailed direct numerical simulations of the SK model on scale-free and Erd\H{o}s-R\'{e}nyi networks for investigating transition to synchronization. Both first order (ES) and second order synchronization transitions are found to occur depending on the values of $\alpha$ and $f$. For SK  model on scale-free networks, we observe that partial degree-frequency correlation enhances the region of existence of ES. On the other hand, SK model on ER networks undergoes mostly second order transition to synchronization for different values of $\alpha$ and $f$. However, we identify small windows of $\alpha$ and $f$ where first order transition is observed.
\section{Acknowledgements}
\par Authors wish to thank C. R. Hens and P. Khanra for insightful comments. P.K. acknowledges support from  DST, India under the DST-INSPIRE scheme (Code: IF140880).

\end{document}